\documentclass[11pt,a4paper]{article}
\usepackage[left=2cm, right=2cm, top=2cm, bottom=2cm]{geometry}

\usepackage{mathptmx} 
\usepackage{hyperref}
\usepackage[parfill]{parskip}
\usepackage{longtable}
\usepackage{graphicx}
\usepackage{authblk}
\usepackage{wrapfig} 
\graphicspath{ {images/} }
\usepackage{units,float,dcolumn}
\usepackage{enumitem,kantlipsum}
\usepackage{setspace}
\usepackage{nomencl}

\makenomenclature

\usepackage[usenames,dvipsnames]{xcolor}
\usepackage[normalem]{ulem} 

\newcommand{\citep}[1]{\cite{#1}}

\usepackage{amssymb}

\usepackage{amsmath,array}		
\newcolumntype{L}[1]{>{\raggedright\arraybackslash}p{#1}}		
\newcolumntype{C}[1]{>{\centering\arraybackslash}p{#1}}	

\usepackage{color}
\usepackage[raggedright]{titlesec} 
\usepackage[font=rm,tableposition=top,format=plain]{caption}		
\usepackage[utf8]{inputenc}
\usepackage[T1]{fontenc}
\usepackage{lmodern}

\makeatletter
\usepackage[backend=bibtex,eprint=false,doi=false,isbn=false,url=false,date=year]{biblatex} 
\bibliography{Voyage2050_HighAngResGWs}

\def\@makechapterhead#1{             
  \vspace*{10pt}                     
  { \parindent 0pt \raggedright
    \ifnum \c@secnumdepth >\m@ne     
      \begin{center}
         \LARGE\sc  \@chapapp{} \thechapter 
      \end{center}
      \vskip 10pt \fi                  
    \begin{center}
      \Huge \bf                        
    #1
    \end{center}
    \nobreak                         
    \vskip 10pt                      
    \rule{\textwidth}{5pt}
    \vskip 20pt                      
  } }

\makeatother

\begin{document}

\newpage
\begin{titlepage}
   \begin{center}
       \vspace*{1cm}
 
       \textbf{\Huge High angular resolution gravitational wave astronomy}
  
       \vspace{3cm}

%
       
\end{center}
\noindent


\noindent
\textbf{Proposing team: } John Baker$^{2}$, 
Tessa Baker$^{3}$,
Carmelita Carbone$^{4}$, 
Giuseppe Congedo$^{5}$, 
Carlo Contaldi$^{6}$, 
Irina Dvorkin$^{1,*}$, 
Jonathan Gair$^{1}$,  
Zoltan Haiman$^{7}$, 
David F. Mota$^{8}$, 
Arianna Renzini$^{6}$, 
Ernst-Jan Buis$^{9}$, 
Giulia Cusin$^{10}$, 
Jose Maria Ezquiaga$^{11}$, 
Guido Mueller$^{12}$, 
Mauro Pieroni$^{13}$, 
John Quenby$^{6}$, 
Angelo Ricciardone$^{14}$, 
Ippocratis D. Saltas$^{15}$, 
Lijing Shao$^{16}$, 
Nicola Tamanini$^{1}$, 
Gianmassimo Tasinato$^{17}$, 
Miguel Zumalac\'{a}rregui$^{18}$

\vspace{0.5cm}
\small{
\noindent
$^{*}$ Contact Scientist, email: irina.dvorkin@aei.mpg.de \\

\noindent
$^{1}$ Albert-Einstein-Institute, Potsdam, Germany \\
$^{2}$ Goddard Space Flight Centre, US \\
$^{3}$ Queen Mary University of London, UK \\
$^{4}$ INAF - Institute of Space Astrophysics and Cosmic Physics, Milano, Italy \\
$^{5}$ Institute for Astronomy, University of Edinburgh, UK \\
$^{6}$ Imperial College London, UK \\
$^{7}$ Columbia University, US \\
$^{8}$ University of Oslo, Norway \\
$^{9}$ TNO, Delft, the Netherlands \\
$^{10}$ University of Oxford, UK \\
$^{11}$ Universidad Aut\'{o}noma de Madrid, Spain \\
$^{12}$ University of Florida, US \\
$^{13}$ IFT, Universidad Aut\'{o}noma de Madrid, Spain \\
$^{14}$ INFN, Sezione di Padova, Italy \\
$^{15}$ CEICO, Institute of Physics of the Czech Academy of Sciences, Prague, Czechia \\
$^{16}$ Kavli Institute for Astronomy and Astrophysics, Peking University, Beijing, China \\
$^{17}$ Swansea University, Swansea, UK \\
$^{18}$ University of California at Berkeley, US
}
       \vfill
       
 \begin{center} 
       A submission to the ESA Voyage 2050 call for White Papers.
 
       \vspace{0.8cm}
  
   \end{center}
\end{titlepage}

\section*{Abstract}
Since the very beginning of astronomy the location of objects on the sky has been a fundamental observational quantity that has been taken for granted. While precise two dimensional positional information is easy to obtain for observations in the electromagnetic spectrum, the positional accuracy of current and near future gravitational wave detectors is limited to between tens and hundreds of square degrees. This lack of precision makes it extremely challenging to identify the host galaxies of gravitational wave events or to confidently detect any electromagnetic counterparts that may be associated with them. Gravitational wave observations provide information on source properties and distances that is complementary to the information in any associated electromagnetic emission and that is very hard to obtain in any other way. Observing systems with multiple messengers thus has scientific potential much greater than the sum of its parts. A gravitational wave detector with higher angular resolution, i.e., better capability to determine the astrometric position of an object on the sky, would significantly increase the prospects for finding the hosts of gravitational wave sources and triggering a multi-messenger follow-up campaign. The angular resolution of gravitational wave detector networks is not fundamentally limited at the degree level. An observatory with arcminute precision or better could be realised within the Voyage 2050 programme by creating a large baseline interferometer array in space. A gravitational wave observatory with arcminute angular resolution would have transformative scientific potential. Precise positional information for standard sirens would enable precision measurements of cosmological parameters and offer new insights on structure formation; a high angular resolution gravitational wave observatory would allow the detection of a stochastic background and resolution of the anisotropies within it; it would also allow the study of accretion processes around black holes and could shed light on the origin of the diffuse neutrino background; and it would have tremendous potential for tests of modified gravity and the discovery of physics beyond the Standard Model. 

\newpage

\nomenclature{AGN}{Active Galactic Nucleus}
\nomenclature{BH}{Black Hole}
\nomenclature{BBH}{Binary Black Hole}
\nomenclature{BNS}{Binary Neutron Star}
\nomenclature{CMB}{Cosmic Microwave Background}
\nomenclature{EM}{Electromagnetic}
\nomenclature{EMRI}{Extreme Mass Ratio Inspiral}
\nomenclature{ET}{Einstein Telescope}
\nomenclature{GW}{Gravitational Wave}
\nomenclature{IMBH}{Intermediate Mass Black Hole}
\nomenclature{IMRI}{Intermediate Mass Ratio Inspiral}
\nomenclature{LIGO}{Laser Interferometer Gravitational-Wave Observatory}
\nomenclature{LISA}{Laser Interferometer Space Antenna}
\nomenclature{LSS}{Large Scale Structure}
\nomenclature{MBH}{Massive Black Hole}
\nomenclature{NS}{Neutron Star}
\nomenclature{SNIa}{Supernova Ia}
\nomenclature{SNR}{Signal-to-Noise Ratio}
\nomenclature{TDE}{Tidal Disruption Event}

\printnomenclature

\newpage
\section{Introduction}
\label{sec:intro}
Over the past few years, the first detections of gravitational waves (GWs) by the ground-based LIGO and Virgo interferometers have had a profound impact on our physical understanding of the Universe. These observations have shed light on the population of compact objects in the Universe and their formation mechanisms~\cite{2018arXiv181112940T}, provided some of the most stringent tests to date of the Theory of General Relativity~\cite{Abbott:2018lct,2019arXiv190304467T} and enabled the first measurement of a cosmological parameter, the Hubble constant, using GW sources~\cite{2017Natur.551...85A}. The impact of GW observations will be further extended by third generation ground based detectors~\cite{punturo2010} and the space-based interferometer LISA~\cite{2017arXiv170200786A}. One fundamental limitation of all of these current and planned detectors is their resolution. GW detectors do not make images and so this is not resolution in the usual astronomical sense, but rather means astrometric precision, i.e., the ability of networks of GW detectors to determine the direction from which the gravitational waves are arriving at the detector. The direction to the first GW sources observed by the LIGO interferometers was only determined to O$(100)$s of square degrees, improving to O$(10)$s of square degrees when Virgo joined the network in 2017~\cite{2018arXiv181112907T}. The anticipated capability of LISA to determine sky location is comparable, being of the order of ten square degrees on average, and one square degree in the best cases~\cite{2017arXiv170200786A}.

High accuracy positional information is relatively easy to obtain for most conventional electromagnetic telescopes and has therefore been a cornerstone of astronomy for centuries. GW observations now and in the near future will not provide accurate positions, but instead give a wealth of alternative information about the intrinsic properties of the sources, such as their masses, rotation rates and distances, that are very hard to obtain electromagnetically. If the GW astrometric positions can be improved, it greatly increases the chance that the GW source can be localised to an individual galaxy or cluster, and that any counterpart emission in the electromagnetic spectrum will be found. This would provide an unprecedented opportunity to understand the relationships between compact object binaries and their hosts, understand the formation channels for these systems and their physical environments and constrain the physical laws driving the emission in the GW and electromagnetic spectra. The observation of the binary neutron star GW170817 by the LIGO/Virgo interferometers~\cite{PhysRevLett.119.161101} has already illustrated the tremendous potential of multi-messenger astronomy. The proximity of that source to the Earth ($\sim40$Mpc) allowed the host galaxy to be determined through the identification of an optical counterpart, which triggered an unprecedented observational campaign with ground and space based facilities~\cite{Abbott_2017_GW170817_MM}. This allowed vastly more to be learnt about the physics of the source than would have been possible using electromagnetic or gravitational wave observations alone.

Astrometric precision can be improved by increasing the baseline in an array of detectors. For ground-based detectors, the size of the network is fundamentally limited by the radius of the Earth, but in space separations that are up to $\sim$AU can be realised. Improvements in angular resolution also come from targeting higher frequencies and improving the sensitivity of the instrument. Ways to increase resolution are discussed in more detail in Section~\ref{sec:strawman}, but an L-class instrument launched by ESA in the Voyage 2050 programme, operating in the millihertz or decihertz GW frequency band and operating in conjunction with a similar instrument, perhaps provided by an international partner, could achieve astrometric precisions at the arcminute level.

Arcminute precision GW astronomy has massive scientific potential.  Identifying the unique host galaxy, or a small number of potential host galaxies to a GW source allows the GW distance information to be combined with electromagnetic redshift information to derive statistical constraints on cosmological parameters and reconstruction of the weak lensing potential (see Section~\ref{sec:cosmology}). Improved astrometric precision also aids better characterisation and subtraction of individual GW sources from the instrumental data, making it more plausible to detect a stochastic gravitational wave background. Missions that are better able to measure astrometric positions will also be better able to constrain anisotropies in any detected background (see Section~\ref{sec:stochastic}). High angular resolution could enable the identification of the hosts of massive black hole (MBH) mergers weeks or even months before the merger, allowing detailed studies of accretion processes in those systems. Additionally, it will allow detailed studies of the hosts of MBH mergers that will shed light on the co-evolution of MBHs with the galaxies in which they reside, and will facilitate searches for neutrinos emitted in these systems. This will be discussed in more detail in Section~\ref{sec:MBH}. Finally, the identification of host galaxies or electromagnetic counterparts to GW sources enable stringent tests of fundamental physics, such as the propagation speed, polarisation state and dispersion of GWs. This is discussed in more detail in Section~\ref{sec:TGR}.


\section{Achieving high angular resolution}
\label{sec:strawman}
Gravitational wave detectors do not directly image gravitational wave point sources, but positional information comes from astrometry. The astrometric precision achievable can be intuitively estimated by scaling the Rayleigh criterion by signal-to-noise ratio (SNR). Thereby $\Delta \theta \sim \lambda/(D \rho)$, where $\lambda$ is the GW wavelength, $\rho$ is the signal-to-noise ratio (SNR) and $D$ is the effective size of the aperture. For a given $\lambda$ we can improve the resolution by increasing $\rho$ and/or $D$. For a single space based detector like LISA the aperture $D$ is synthesised by the motion of the detector around the Sun and can be as large as $\sim$AU for sources that are long-lived. For sources that are short-lived, for example massive black hole mergers where the majority of the SNR is accumulated in the final week before merger, the effective aperture is much smaller.

A large aperture can be synthesised even for short-lived sources by having more than one detector operating simultaneously with a separation of a significant fraction of an AU, for example with one detector leading the Earth and one trailing the Earth in its orbit. While we might expect some economies of scale, it is likely that two LISA-like detectors would not be affordable within the L-class ESA mission budget. However, several international partners have advanced plans or the expertise to launch a space-based interferometer on the Voyage 2050 timescale, including China~\cite{2016CQGra..33c5010L}, Japan~\cite{Sato_2017} and the US. High angular resolution could therefore be achieved by an ESA launch of a single space-based interferometer to coincide with data taking of these other instruments. Such a mission would be L-class, but not smaller.

Improved angular resolution also comes from increased SNR and from shorter wavelengths. Therefore, detectors with improved sensitivity operating at higher frequency would also provide increases in resolution. In LISA Pathfinder the acceleration noise at these frequencies was dominated mostly by gas pressure which was orders of magnitude higher than the pressure that is routinely achieved in UHV chambers. Other limiting noise sources can be reduced by using larger gaps, better -- maybe even active -- gravitational balancing, $\mu$N thrusters with a faster response time to reduce residual spacecraft motion, and maybe an optical readout system which monitors all degrees of freedom of relative spacecraft to test mass motion to improve better calibration and subtraction of these forces and torques. Based on the performance of LISA Pathfinder and the projected performance of LISA, these improvements should allow the acceleration noise to be reduced significantly at higher frequencies. Operating at frequencies around $1$ decihertz requires a constellation with smaller inter-spacecraft separations, which for the same optical parameters, would allow to reach LISA's shot noise limit in terms of strain sensitivity at higher frequencies. Further shot noise reductions require higher laser power ($\propto \sqrt{P}$) and larger telescope diameters ($\propto D^2$~). Beyond this, significant improvements in phasemeter technology and the timing and ranging system as well as ways to minimize, for example, tilt to length coupling will be required to reach shot noise limited performance at this new level. 

Such a detector concept, ALIA, was considered in~\cite{2005PhRvD..72h3005C}, along with several other ideas for constellations of gravitational wave interferometers. The ALIA concept assumes acceleration noise a factor of $10$ lower than the LISA requirement, and positional noise a factor of $100$ better. Such improvements would require research and development but are certainly achievable within the Voyage 2050 timeframe. A network of two detectors with the sensitivity of ALIA would typically localise sources to arcminute precision, which is two orders of magnitude better than LISA. We note that one ALIA-like instrument already has exciting scientific potential, particularly working in conjunction with future ground-based interferometers. This science is described in a separate submission to the Voyage 2050 call, ``The missing link in gravitational-wave astronomy: Discoveries waiting in the decihertz range'', and is not discussed here. We focus on what is achievable with high angular resolution. A network of one space-based and one ground-based interferometer will also provide an $\sim$AU baseline for events that can be observed simultaneously with high SNR, such as the mergers of binaries of intermediate ($\sim 100$--$1000M_\odot$) mass black holes. Sources that are observed at different times from the ground and in space will typically be slowly evolving in the space-based detector band and therefore already have a large synthesised aperture before the observation by the ground-based detector. For such systems the gain in resolution is primarily driven by increased signal-to-noise and is limited by the precision with which the arrival time at the second detector can be predicted by the observation with the first.

In this proposal we discuss what could be achieved with arcminute angular resolution and do not make detailed reference to a mission concept
. The preceding discussion illustrates that such a resolution is in principle achievable for an L-class mission on the Voyage 2050 timescale, although it would almost certainly rely on collaboration with an international partner. Angular resolution as small as arcseconds could be achieved by multi-interferometer concepts such as the Big Bang Observer~\cite{2005PhRvD..72h3005C}. Such a mission is most likely unfeasible for ESA alone on the Voyage 2050 timescale, but some results for that resolution will also be given for illustration.

\section{Science with high angular resolution gravitational wave observations}

\subsection{Statistical cosmology}
\label{sec:cosmology}

Cosmology is currently facing two unresolved ``tensions''. The first concerns the $H_0$ parameter that has been measured to a precision of a few percent. 
The best constraints we now have are discrepant at the level of 4.4$\sigma$: CMB angular 
diameter distance measurements at $z\sim1,100$ \cite{planck2018}, and SNIa surveys via 
luminosity distance measurements at $z<1$ \cite{riess2019}. More recent results have 
been obtained by calibrating distances using the Tip of the Red Giant Branch 
\cite{freedman2019} instead of the Cepheids, which leads to a measured $H_0$ midway 
between the values from CMB and SNIa. Contrarily, an independent measurement using lensed quasars 
has found that the discrepancy with CMB is even bigger, at the 5.3$\sigma$ level 
\cite{wong2019}. 

The second tension happens purely in the dark matter sector defined by the 
$\Omega_m$-$\sigma_8$ parameter space, where $\Omega_m$ is the dark matter density parameter and $\sigma_8$ 
is the rms of the matter fluctuations (amplitude of linear matter power spectrum) 
at a scale of 8$h^{-1}\unit{Mpc}$. A recent reanalysis of both the Kilo Degree Survey and the 
Dark Energy Survey has confirmed full consistency between the two experiments, yet 
again a tension of 2.5$\sigma$ with the CMB measurements \cite{joudaki2019}.

GW astronomy has provided us with a new way to do cosmological measurements that 
are completely independent from electromagnetic (EM) observations. For coalescing binary systems, the strain is proportional to the redshifted chirp mass to the 
power of 5/3, and inversely 
proportional to the luminosity distance, $d_L$. Therefore, these two quantities can be jointly inferred directly from the GW 
inspiral+merger signal by measuring the amplitude and frequency evolution over 
time. With a single detection of a binary neutron star (BNS) merger, GW170817 -- a bright standard 
siren so called because it had a multitude of EM follow-up observations -- $H_0$ has been 
constrained to $\sim$15\% precision level \cite{2017Natur.551...85A,Fishbach:2018gjp}, with percent level achievable in a few years time by combining future similar observations \cite{Chen:2017rfc}. Even without any EM counterpart, the GW170814 dark standard siren event 
brought a measurement of $H_0$ with a 48\% precision \cite{Soares-Santos:2019irc}, thanks to the synergy with galaxy surveys that provided a 
calibrated galaxy sample selected in the GW position error box for statistical inference of the redshift.

With the upcoming third generation ground-based detectors, such as the Einstein Telescope (ET) 
\cite{punturo2010}, and the space-based detector LISA \cite{2017arXiv170200786A}, GW 
cosmology will be taken to a whole new level. These detectors will reach a median horizon of $z\sim1$ from the ground and $z\sim2$ in space, hence 
improving statistical power to constrain other cosmological parameters, such as 
those for dark energy \cite{2016JCAP...04..002T,tamanini2017,Caprini:2016qxs,Cai:2017yww} or modified gravity \cite{belgacem2019}. Also, the position 
error box will shrink down to a few deg$^2$ or better, which will improve the chance of identifying an EM counterpart. Finally, the 
event rate will dramatically increase, reaching a few thousand sources by the 
2030s, significantly improving the statistics.

A detector working in the decihertz band with $\sim\unit{arcmin}$ resolution (or better) will be 
capable to revolutionise cosmology as we know it today, even after Euclid 
\cite{laureijs2011} and the Large Synoptic Survey Telescope (LSST) \cite{2009arXiv0912.0201L} for the following reasons:
\begin{itemize}[wide, labelwidth=!, labelindent=0pt]
\item An angular resolution of $\lesssim\unit[1]{arcmin^2}$ will make the identification of the EM 
counterpart almost guaranteed for most of the GW sources. Euclid and LSST will 
provide galaxy catalogues for $\sim40\%$ of the entire sky, so the galaxy number 
density will be, on average, $\unit[30]{arcmin^{-2}}$ at the Euclid depth (slightly higher for 
LSST), hence making the identification of the host galaxy relatively 
straightforward. Likewise for any spectroscopic follow-up, with marginal redshift 
errors. 
\item Typical sources in the decihertz will be persistent, hence allowing high SNR 
detection with $d_L$ measured to 1\% precision or better, likely to 0.1\% over a broad range of 
redshifts, which will provide great leverage on the distance vs redshift relation 
to jointly constrain a number of cosmological parameters. 
\item Multi-band GW observations, where decihertz sources are subsequently observed by the network of 
ground-based GW detectors operating above $\unit[1]{Hz}$ will allow joint inference of 
source parameters, including $d_L$. 
\item High angular resolution attained by cross-correlating over a long baseline 
will allow accurate determination of the polarisation angle, hence breaking the
degeneracy between distance and inclination, which was a major source of uncertainty for GW170817.
\end{itemize}

In the decihertz frequency band there are mainly two classes of GW 
sources that can be exploited for cosmology: stellar-mass binary black 
holes (BBHs) and binary neutron stars (BNSs). For both of them a space-based decihertz detector will observe the long-lasting inspiral phase, 
extracting accurate measurements of the binary's parameters. The high 
angular resolution will further boost the cosmological potential of 
these detections. BBHs and BNSs are, however, expected to contribute 
differently.

Stellar-mass BBHs will be detected roughly at the same rate as current LIGO/Virgo sources, 
since no long-living BBH should appear in the decihertz range. The total 
number of detections will depend on the sensitivity of the instrument. 
Achieving a localization of $\sim\unit{arcmin}$ or better will provide 
accurate forewarnings to Earth-based detectors, and will drastically 
reduce the localization volume. Although no EM 
counterparts are expected for BBHs, the small localization volume will 
contain a reduced number of possible host galaxies, leading 
to the identification of a single host galaxy in the most favorable 
cases. By taking results from the proposed Big Bang Observer telescope (BBO; see Fig.~6 of \cite{cutler2009}) and 
simply degrading the angular resolution from $\unit{arcsec^2}$ to $
\unit{arcmin^2}$ (and assuming roughly the same distance measurements), 
we find a maximum of few tens/hundreds of galaxies within the volume 
error box at each redshift, and the identification of the unique 
hosting galaxy for all BBHs below redshift 0.5. This will improve the 
constraining power of the ``statistical'' cosmological analyses 
\cite{schutz1986,DelPozzo:2011yh}, providing an independent test not 
only of $H_0$, but also of other cosmological parameters. These 
considerations will be particularly relevant for BBHs jointly observed 
both in space and on Earth by third generation detectors such as ET or the Cosmic Explorer (CE), whose 
investigation can be improved and pushed to higher redshift with a 
space-based interferometer in the decihertz frequency band.

The most promising sources for cosmology in the decihertz band are, 
however, BNSs, especially with a coincident Earth-based detection. In this case the excellent sky localization achieved long in 
advance of the merger, will give enough time to detect the associated EM signal (kilonova and possibly a gamma-ray burst) for all the detected BNSs, which 
can then be used as bright standard sirens for cosmology. The actual rates 
and horizon of BNSs will depend on the sensitivity of the instrument, 
but drawing from similar results for DECIGO \cite{cutler2009,Nishizawa:2010xx}
one can comfortably assume that a Hubble diagram with more than $
\sim10^5$ events out to redshift $\sim$3 can be constructed, 
especially if joint detections with Earth-based detectors will be 
possible. With these numbers one can hope to obtain sub-percent 
constraints on $H_0$ and also to probe the equation of state of dark 
energy at the 10\% level or better \cite{cutler2009}. 

\begin{wrapfigure}{r}{0.6\textwidth}
\centering
\includegraphics[width=0.99\linewidth]{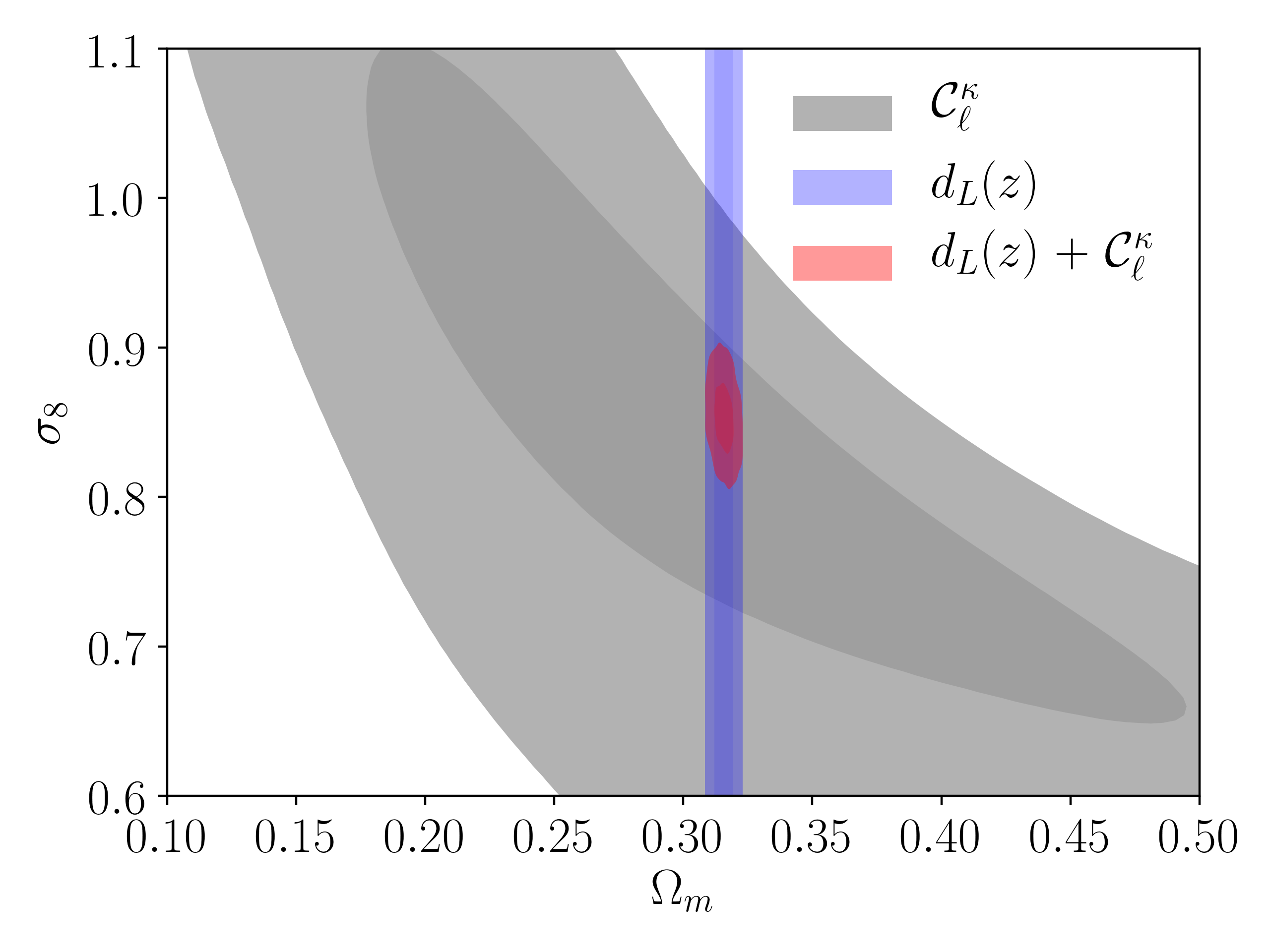} \\
\includegraphics[width=0.99\linewidth]{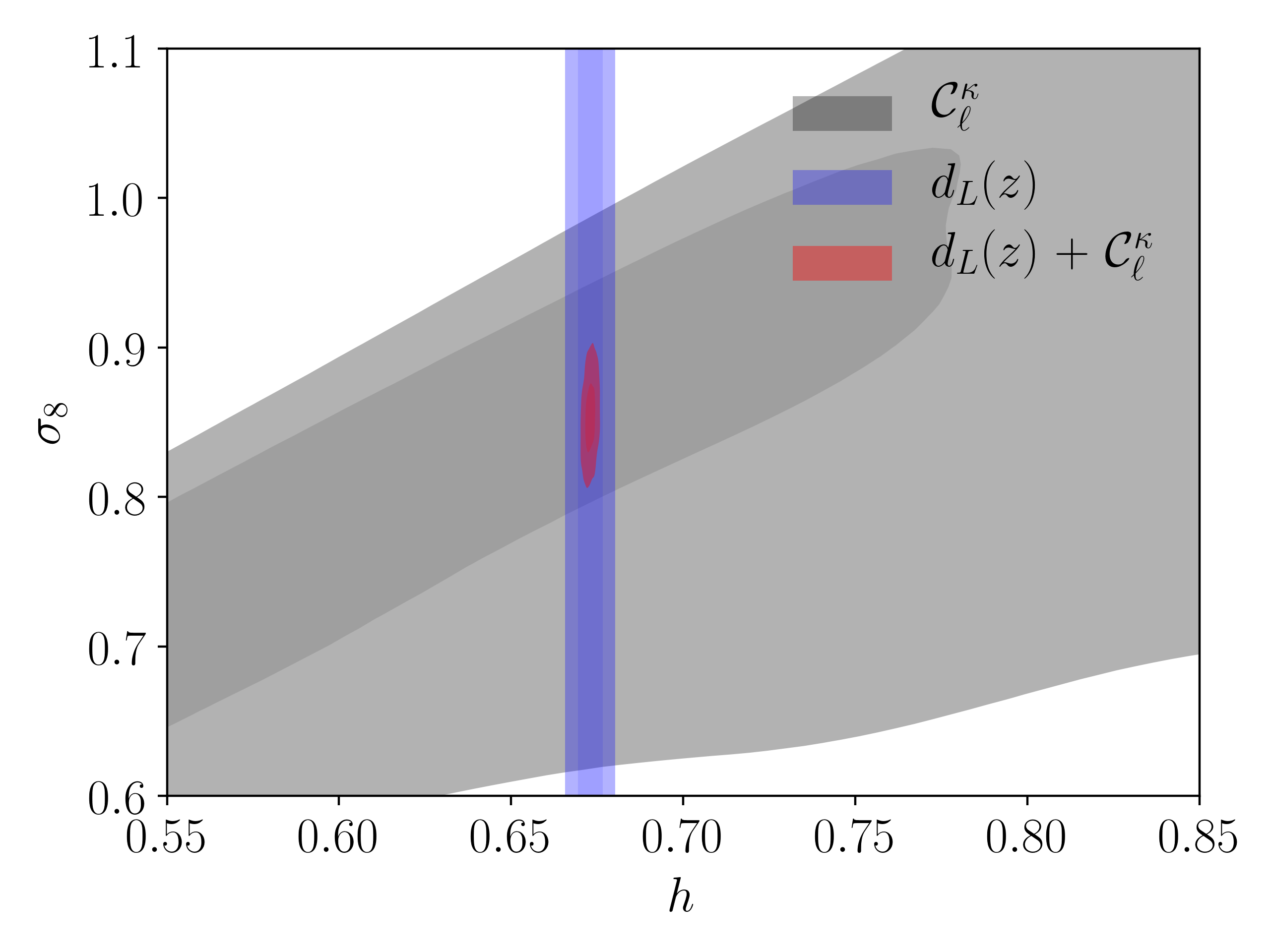}
\caption{Adapted from Fig.\;1 of Ref.\;\cite{congedo2019}, cosmological forecast for the flat $\Lambda$CDM model
assuming a very conservative source number density of $\unit[1]{deg^{-2}}$.
The figure shows the constraints from luminosity distance only, $d_L$, the
weak lensing convergence power spectrum, $\mathcal{C}^\kappa_\ell$, and jointly.
Decihertz arcmin resolution would guarantee up to a factor of 50 improvement in number density (likewise for shot noise),
hence allowing access to non-linear scales.
The constraining power would be improved almost surely by an order of magnitude.}\label{fig:cosmology}
\end{wrapfigure}

Furthermore by 
the 2035-2050 time period we will also have a good handle on the 
physics of neutron stars and might be able to use this information to 
estimate the redshift of the event directly from the GW signal \cite{Taylor:2011fs,DelPozzo:2015bna}.
This is another reason to argue that each BNS will have an associated 
redshift measurement, and that a Hubble diagram with $10^5 - 10^6$ 
standard sirens could be built.

In the decihertz band one should also be able to see the merger of 
intermediate-mass BBHs, although the rates so far are unknown. Moreover it is not as yet clear if any EM 
counterpart signal could be observed for these events. Given these 
uncertainties, it is impossible to predict if intermediate mass BBHs 
will be useful in a cosmological context, although their likely 
larger horizon makes them an appealing class of GW standard sirens.

Besides probing the cosmic background expansion through the distance-redshift relation \cite{2016JCAP...04..002T,tamanini2017}, high angular resolution GW astronomy also yields unprecedented accuracy to 
constrain two key properties of our universe: geometry and dark matter clustering.
This will be attained by additional observables as recently illustrated by \cite{congedo2019} -- all measured by a single detector:

\begin{itemize}[wide, labelwidth=!, labelindent=0pt]
\item {\bf The weak lensing power spectrum}, derived through statistical analysis of rms 
fluctuations around the nominal distance vs redshift relation \cite{cutler2009}, 
which is sensitive to dark matter clustering through $A_s$, $n_s$, and $\sigma_8$ as shown in Ref.\;\cite{congedo2019}. Given 
the negligible measurement error on $d_L$, the rms error is itself a point estimate 
of convergence at the source position. The statistical power obtained by a high 
angular resolution GW detector would be at least a factor $10^3$ better than galaxy 
lensing surveys, hence allowing very precise constrains in the $\Omega_m$-$\sigma_8$ space
\cite{congedo2019}, and also on extensions to the $\Lambda$CDM model \cite{Camera:2013xfa}.
As shown in Ref.\;\cite{congedo2019}, the constraining power on all cosmological parameters would be enhanced if a joint analysis of distance-redshift relation and lensing were to be carried out by such a high resolution decihertz detector.
In fact, given an expected total number of events between $10^5$--$10^6$, a ballpark estimate based on Ref.\;\cite{congedo2019} guarantees a source number density between $\unit[5-50]{deg^{-2}}$,
which would be more than enough to beat down shot noise and access the non-linear scales of $\ell\sim1,000$ or above.
This would allow us to comfortably reach the $0.1\%$ level on all cosmological parameters,
that would significantly outperform galaxy lensing surveys of the future.
An example of the cosmological results that could be achieved with
such high angular resolution is illustrated by the forecast of Fig.\;\ref{fig:cosmology}, which shows the constraints on the rms of matter fluctuations $\sigma_8$, the dark matter density parameter $\Omega_m$ (upper panel) and the Hubble parameter $h$ with $H_0=100h$ km/s/Mpc (lower panel).
\end{itemize}

\begin{itemize}[wide, labelwidth=!, labelindent=0pt]
\item {\bf The one-point probability density function of lensing convergence} (with the aid 
of cosmological simulations) is also sensitive to clustering \cite{patton2017}.
A high angular resolution detector will be able to reconstruct the pdf of the measured convergence field,
and allow the extraction of additional information.
This is a probe that would complement luminosity distance and weak lensing \cite{congedo2019}.
\item {\bf Gravitational lensing} also affects the amplitude and polarisation of GW 
backgrounds. In fact, GWs follow null geodesics of the space-time and their  
propagation suffers distortions due to the large scale structure (LSS) 
distribution in the Universe as happens to photons in the geometric optics 
approximation. In addition, as in the photon case, effects such as the Shapiro time 
delay and the Integrated Sachs-Wolfe (ISW) due to the time evolution of the 
gravitational potentials in the presence of dark energy or massive neutrinos 
affect GW propagation \cite{Contaldi:2016koz,Cusin:2017fwz, Bertacca_etal_2017}. Modelling 
unperturbed GW backgrounds and comparing with future high angular resolution 
observations (affected by cosmological perturbations) in GW total intensity and 
polarisation will provide the possibility to extract information about the 
underlying dark matter distribution as can be performed in the case, for example, of CMB-
lensing reconstruction (see also Section \ref{sec:agwb}).
\item {\bf Non-linear effects induced in the GW propagation by large scale structure} correlate with 
other cosmological probes such as galaxy-lensing and galaxy clustering, which trace 
the same underlying dark matter distribution. Such cross-correlations represent 
further probes to be exploited in the determination of cosmological parameters 
affecting the LSS distribution. On the other hand, the non-linear evolution of 
structure formation needs to be taken fully into account in the accurate 
modelling of such new signals, especially in the presence of modifications to the 
standard cosmological framework. This will be achieved in the near future via the 
use of full-sky simulations of GW maps accounting for the non-linear structure 
evolution as provided by large cosmological simulations in different cosmological 
scenarios \cite{Jenkins_etal_2018,2018JCAP...09..039S,Cusin:2018rsq,Cusin:2019jpv}. Such modelling will soon improve estimations 
of the luminosity-distance relation of GW sources, as well as modifications in GW 
intensity and polarisation due to the evolution of intervening structures during 
their propagation.
\item {\bf Cross-correlations of the GW signal with future surveys}, such as Euclid and SKA, 
will also make possible the so-called GW delensing. This will allow  the full constraining power of future high angular 
resolution GW experiments to be exploited in extracting cosmological parameters.  On the other 
hand, data delensing could be achieved also via mock simulated weak-lensing maps 
from cosmological simulations. This approach will also help to break 
degeneracies between non-standard cosmology and GW lensing effects.
\end{itemize}

\subsection{Stochastic background detection and characterisation}
\label{sec:stochastic}


The stochastic background is, by definition, made up of an incoherent
superposition of signals from multiple sources that are unresolved in
both the time and angular domain. A stochastic background will exist
for most sources being probed by gravitational wave detectors if the
distribution of luminosity of the underlying population extends below
the detection threshold of the observatory or if the signal duration
and event rates are such that an incoherent superposition is
guaranteed. 
In the millihertz to decihertz frequency range this will be the
case for galactic and extra-galactic compact binary systems~\cite{2018CQGra..35e4004M,Regimbau:2016ike, Dvorkin:2016okx, Nakazato:2016nkj, Dvorkin:2016wac, Evangelista:2014oba}.

Another class of stochastic backgrounds is that generated by phase
transitions in the early universe \cite{2007PhRvL..98k1101S,2000PhRvL..85.3761D} or perhaps a
gravitational wave background generated during the epoch of
reheating. Although the mechanism is very different these will also be
incoherent given that the horizon scale at the time of generation is tiny
compared to today.

Yet another class is the primordial background due to
inflation \cite{1997PhRvD..55..435T,1979JETPL..30..682S}. During the epoch of inflation, tensor modes are inflated to
super-horizon scales where their comoving amplitude becomes
constant. Once they re-enter the horizon, during the deceleration
phase of the expansion, these modes will appear as squeezed
gravitational waves. The squeezing results in phase correlations which
are possible only in modes that have re-entered the horizon. The phase
correlations are a distinguishing feature of a primordial inflationary
background although, on observable scales/frequencies, the correlation
will be destroyed by sub-horizon metric perturbations. As such, a
primordial background can also be thought of as an incoherent one for
observational purposes. A primordial background is expected to be the
smallest in amplitude by some orders of magnitude with $\Omega_{\rm gw} \sim 10^{-15}$ for inflationary potentials on the order of
$10^{16}$ GeV, where $\Omega_{\rm gw}$ is the energy density in GW per unit logarithmic frequency in units of the critical density of the Universe.

Astrophysical and cosmological information is contained in both the
average (monopole) amplitude $\Omega_{\rm gw}$ and anisotropies of any
background. The amplitude of the background is an integrated measure
of the underlying population which probes very different limits of the
distribution than the collection of single, high signal-to-noise,
detections. The anisotropies contain information about the angular
distribution of the sources. This information can be used either as an
aide to source separation in conjunction with spectral resolution or
as a tracer of astrophysical or cosmological structure. 

\subsubsection{Constraining backgrounds and their anisotropies}

The angular resolution of a GW detector depends on its configuration as well as on the type of source (see Section \ref{sec:strawman}). In particular, if we consider astrophysical resolvable sources, a space-based detector, like LISA, is not a pointed instrument but an ‘all-sky monitor’. Ground-based detectors share the same property, but since there is a network of such detectors, the signals can be correlated. 
This method cannot be extended to a single space-based detector. Instead, the motion of the satellite must be considered and eventually the combination of the two time series that can be extracted from a single detector, or correlating with a future ground-based detector working in the same frequency range. 
The situation is different for stochastic gravitational wave backgrounds, both of astrophysical and cosmological origin, where the duration of the signal is very long (infinite) compared to the time of observation. Given that the noise in different detectors is uncorrelated, while the signal is expected to be correlated, a way to circumvent this problem and to improve the signal-to-noise-ratio is to cross-correlate the output from two detectors.

We can therefore distinguish between two types of angular
resolution afforded by GW detectors. The first is the
astrometric resolution - this is the resolution with which the
instrument can pinpoint the location of a single coherent source. To achieve the quoted
level of $\sim$arcmin the full phase information of the
coherently observed signal is used.

For incoherent signals this resolution limit is not achievable. In
this case angular resolution is limited by the combination of the
detector response functions and the baseline over which the
cross-correlation of individual signals is being carried out. The
angular response function of any detector is severely limited by the
fact that no focusing of gravitational waves can be achieved on
practical scales. The baseline and frequency coverage are therefore
the most important factors that determine the angular resolution of
any set up.

Astrometric resolution however is still of fundamental importance in
searching for backgrounds. The efficiency with which individual
signals can be identified and subtracted (masked) in the time domain is
correlated with the phase sensitive, astrometric resolution. In fact,
in order to perform a detection and possibly, a characterization, of
any cosmological backgrounds, the subtraction of galactic and
extragalactic foregrounds/backgrounds must be performed with extremely
high precision. For this purpose, it is crucial to exploit the fact
that different information can be extracted by performing
cross-correlation of the data sets in time and/or frequency
domains. For example, while cosmological backgrounds are expected to
be stationary, the foreground due to the unresolved white dwarf
binaries in the galaxy is expected to present some yearly
modulation~\cite{2014PhRvD..89b2001A} which can be used for component
separation. Analogously, since the spectral shapes of backgrounds
arising from different sources are not expected to match, different
signals can be disentangled by accurately modeling the different
components. It is worth stressing that while in very small frequency
ranges different signals may appear to be degenerate, this degeneracy
is eventually broken for sufficiently large intervals. 

\begin{figure}[t]
\centering
\includegraphics[width=.9\linewidth]{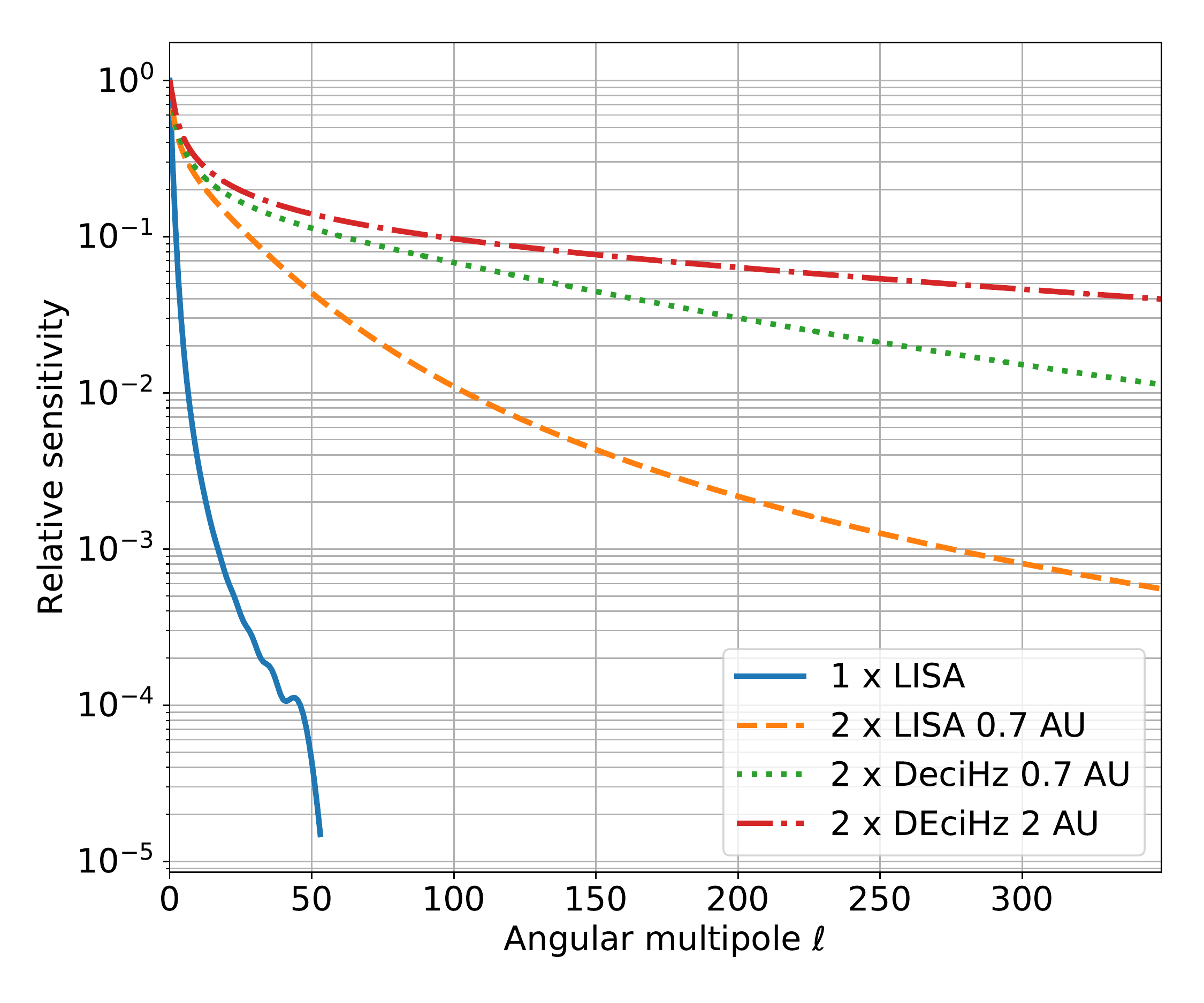}
\caption{Idealised angular resolution for different detector configurations (see text). The relative sensitivity to different angular multipoles is obtained by integrating the spherical response to a cross-correlation baseline weighted by a simple model of the noise spectrum based on the individual detector arm lengths. Convolution with the detector response functions and the sky-phase coverage would give additional structure on top of the idealised case.}\label{fig:angular}
\end{figure}

The ideal configuration is therefore one that maximises the astrometric resolution in the given frequency range \emph{and} the 
angular resolution so that the unresolved signal can also be masked or cross-correlated with other tracers most efficiently. 
This can be achieved with an optimised configuration of detectors whose signal can be cross-correlated over large distances. An idealised figure of merit for the angular resolution can be obtained by integrating
the contribution of each frequency to spherical multipoles on the sky. The plane waves at each frequency can be expanded
in spherical Bessel functions $j_{\ell}$ to obtain an angular response
\begin{equation}
 {\cal A}_\ell = \int^{f_{\rm max}}_{f_{\rm min}} \, df \, w(f)\, j_\ell\left(2\pi \frac{f b}{c}\right)\,,
\end{equation}
where $b$ is the length of the baseline formed by the cross-correlation of detector signals and $w(f)$ is a weighting
function determined by the high-frequency noise of the detectors. This figure of merit is idealised and the actual
response would be a convolution of the full spherical mode expansion with the detector response function and sky modes
given a particular phase coverage on the sky but it gives a limiting case for relative comparisons between configurations.

Figure~\ref{fig:angular} shows the normalised response function for four different configurations: single LISA, 2 LISA-type detectors with a $0.7$ AU separation, 2 decihertz detectors with a $0.7$ AU separation and 2 decihertz detectors with a $2$ AU separation. For the decihertz detector case we used an  ALIA-like configuration as a prototype~\cite{2005PhRvD..72h3005C}. The dominant contribution to angular resolution is given by the longest cross-correlation baseline length. For a single LISA this is the same order of magnitude as the individual arm lengths. However, having two LISA-type detectors dramatically increases the angular resolution. The optimal configuration consists of two detectors separated
by a distance of the order of an AU with sensitivity peaking in the decihertz range, which would have an angular resolution for stochastic backgrounds of around a degree. This resolution would greatly facilitate the separation of galactic and extra-galactic stochastic signals and also
enable the search for statistical, cosmological effects by cross-correlation with large scale structure \cite{Contaldi:2016koz,Cusin:2017fwz,Cusin:2018rsq,Cusin:2019jpv}.

\subsubsection{Stochastic background anisotropies from astrophysical and cosmological sources}
\label{sec:agwb}

The anisotropies of the stochastic background are a unique observable that contains both astrophysical and cosmological information. Astrophysical sources that contribute to the stochastic background in the millihertz-decihertz frequency range (stellar-mass compact binaries, intermediate-mass BH binaries, extreme mass ratio inspirals) reside in galaxies, and it is therefore expected that the intensity of the background will depend on sky direction, analogously to the cosmic infrared background. As shown in \citep{Cusin:2019jpv,Cusin:2019jhg}, both the amplitude and the shape of astrophysical component of the stochastic background anisotropies depend on the formation and evolution processes of binary compact objects (such as the properties of their stellar progenitors, supernova explosion mechanism etc.). Crucially, the detection of individual merging binaries will only provide information on the brightest sources in the population, in contrast to the stochastic background. Detecting the anisotropies of this background will allow us to study how the properties of GW sources in the faint end of the distribution correlate with those of their host galaxies. For example, a signal from a population of primordial BHs, formed in the early Universe and presumably closely following the distribution of dark matter, would have different angular power spectrum compared with the signal from stellar-origin BHs which form in the late Universe and reside in luminous galaxies, which are a biased tracer of the dark matter distribution~\cite{2018JCAP...09..039S}.

The latest observational upper limits from  the first and second Advanced LIGO runs~\cite{LIGOScientific:2019gaw} are derived for multipoles only up to $\ell=4$ and are several orders of magnitude above current theoretical predictions \citep{Cusin:2018rsq,Jenkins_etal_2018,Cusin:2019jpv,Cusin:2019jhg}. Major advances both in detector sensitivity and in analysis methods are required in order to be able to detect the anisotropic component of the stochastic background. Moreover, since the astrophysical background is expected to dominate any cosmological backgrounds from the early Universe, it is necessary to detect and fully characterize the former in order to be able to also measure the latter.

One of the limiting factors in observing the stochastic background anisotropies in the $10-100$ Hz frequency band, accessible to ground-based detectors, is the time-like shot noise, which arises because the signals from merging binary compact objects have a very short duration with respect to the integration time and almost no time overlap, especially in the case of BBHs. As a result, shot noise is expected to dominate the signal for any realistic time of integration \citep{Jenkins:2019uzp,Cusin:2019jpv,Jenkins:2019nks}. 
However, in the decihertz frequency range the astrophysical background can be considered stationary, since each individual signal duration (the inspiraling phase of a binary compact merger) is longer than the time of observation.

Therefore, the best observational setup that will allow the detection and characterisation of the stochastic background anisotropies is a combination of two detectors operating in the millihertz to decihertz frequency range, which will alleviate the shot noise problem and allow small angular scales to be resolved (see Figure \ref{fig:angular}).


Finally, as mentioned in Section \ref{sec:cosmology}, GWs are distorted by intervening LSS and are affected by the Shapiro time delay and ISW effect, similarly to photons. These effects are imprinted in the angular power spectrum of the stochastic background and can be used to study the LSS~\cite{Cusin:2017fwz,Cusin:2018rsq,Cusin:2019jpv}. Moreover, the astrophysical GW background is expected to be a biased tracer of the galaxy distribution. A particularly promising way to study these effects is to cross-correlate the GW map with EM observables, such as the weak lensing map or galaxy number counts. EM-GW cross-correlation can also help to extract the astrophysical signal in the presence of a shot noise dominated background map.

Stochastic backgrounds are also produced by sources in the early Universe. For example, the background produced by a cosmic string network will have an anisotropic component that depends on the string tension~\cite{Jenkins:2018lvb}. In general, the plethora of stochastic backgrounds contributing in the same frequency band may make it difficult to distinguish them via the observation of the monopole alone. Observing the anisotropies will help in identifying the various contributions to the overall stochastic signal.

\subsection{Multi-messenger observations of massive black holes}
\label{sec:MBH}
Massive black holes (MBHs), ubiquitous in galactic nuclei, play an important role in galaxy evolution. Binary MBHs, expected to form following galaxy mergers, are prime GW sources, targeted by LISA and pulsar timing array experiments.  
Multi-messenger observations of binary MBH mergers are the key to understanding how they form and co-evolve with their host galaxies. In particular, high angular resolution GW observations will allow the localization of binaries well in advance of the merger (a few months or more), so that electromagnetic observatories can search for characteristic merger signatures. These observations will provide unique information on the properties of accretion disks that fuel active galactic nuclei (AGNs). Moreover, coincident neutrino-GW observations that may be possible with a high-resolution GW detector, will allow the identification of the sources of the diffuse neutrino flux and to study particle acceleration processes.

\subsubsection{Accretion disks of active galactic nuclei}
\label{sec:agn}
Despite the fact that quasars
and AGN are a key ingredient in galaxy formation, our
knowledge of the basic properties of the
accretion disks that fuel them, such as their density, temperature, geometry, accretion rate or
lifetime, remain poorly understood~\citep{Martini01}. GW signatures of
mergers involving MBHs and/or stellar-mass BHs in an AGN disk can
complement EM data and provide novel information on disk properties
by probing beneath the AGN photosphere~\cite{Ford19,Inayoshi:2017hgw,Tamanini:2019usx}.

First, with unique identifications of host galaxies of MBH
mergers we will, for the first time, have direct,
detailed data on the bright EM emission (e.g. spectrum, light-curve)
from MBHs whose masses, spins, and orbital parameters are precisely
known, opening up new ways to study the physics of accretion.

Second, a high-resolution GW detector will be able to directly measure the effect of gas drag
on the GW waveform of stellar-mass BHs or intermediate-mass BHs
merging with the MBH or in intermediate mass-ratio inspirals (IMRIs).  The imprint of gas drag on the GWs will reveal 
average AGN disk properties underneath the EM photosphere \citep{Yunes_2011,Kocsis_2011,McK14,LISA-IMRI-gas}.
The rate of gas hardening of BH binaries implicitly reveals the relative
importance of dynamical hardening in the AGN channel of BH mergers.
If gas hardening is too slow or inefficient, tertiary encounters within the disk
are required to harden a binary to merger \citep{Leigh18}.
The magnitude and frequency-dependence of the deviation from vacuum waveforms will probe
disk properties and with high signal to noise measurements they can be disentangled from uncertainties in MBH
binary parameters~\citep{Kocsis_2011,Yunes_2011,LISA-IMRI-gas}.

Third, we will be able to detect IMRIs and identify them with known AGN. Identifying IMRIs in AGNs will
constrain both the structure and the lifetime of AGN disks by implying the existence of migration traps
and information on the timescale for migration within the disk. Indeed, depending on AGN disk structure,
gas torques cause embedded migrators
to converge at traps where large mass intermediate-mass black holes (IMBHs) can be built up via hierarchical
mergers \citep{Secunda19,McK19a,Yang19}. If AGN episodes persist long enough ($\geq 1$Myr),
most AGNs should produce IMBHs, yielding a large population of MBH-IMBH binaries and IMRIs \citep{McK12,McK14, McK19a}. 

Finally, GW observations are complemented by EM observations of
broad emission lines in both X-ray and optical bands.  The
presence of a binary strongly perturbs the nearby gas disk and 
changes the kinematics of emission lines and imprints unusual
periodic variability. Binary stellar-mass BHs or IMRIs resolved  in GWs are precursors to
EM signatures that can outshine moderate luminosity AGN disks,
if due to Hill sphere implosion \citep{McK19b}. EM effects can be especially
large in X-rays, probing gas closest to MBHs.  The broad Fe K$_{\alpha}$ line,
and other broad X-rays lines detectable with \emph{Athena} can display strongly disturbed spectral
shapes, and/or periodic Doppler modulations on the binary's orbital
timescale of O(hr)~\citep{McK13,Sesana+2012,McK15}.  With simultaneous
EM and GW observations these effects will be especially robust, and will strongly constrain system parameters (e.g. disk
properties, as well as BH spins and masses). Moreover, the absence of EM counterparts
to resolved GW emission from an AGN will allow us to strongly constrain accretion disk optical depths.

\textbf{Precursor observations}

The EM counterpart of the MBH binary merger needs to be identified to
accomplish the above scientific goals.  This is likely to be hampered by the lack
of an ab-initio understanding of binary accretion and the
corresponding spectral evolution properties.  This necessitates
sufficiently accurate GW localization to allow for the host of the GW source to be
uniquely identified.

For much of the science above, it will be necessary to identify the GW
source well in advance of the merger, so that the GW chirp and any
corresponding EM chirp can be observed in tandem, for at least a few
hundred cycles.

A promising signal is a quasi-periodic EM ``chirp'', tracking the
phase of the GWs. The torques from the binary are expected to create a central cavity in
the surrounding disk, nearly devoid of gas, within a region about
twice the orbital separation \citep{al94}.  However, numerical
simulations have found copious gas inflow into this cavity
\citep{Roedig+11,ShiKrolik2012,Noble+12,Dorazio+2013,Farris+2014a},
fueling accretion onto the BHs and producing detectable EM emission.
Because of copious shock-heating, gas near the BHs in this late stage
is expected to be hotter than in the case of a single-BH AGN
\citep{ShiKrolik2012,Roedig+2014,Farris+2015a,Bowen+2017,Tang+2018}.
The corresponding UV/X-ray emission would have different (harder)
spectra, with possible signatures of a disk cavity in the form of a
'notch'~\citep{Roedig+2014}.

Most importantly and robustly, the EM light-curve should display a
characteristic modulation on the orbital timescale of an $\sim$hour (a
month before the merger) to $\sim$ minutes (a day before the merger).
Hydrodynamical simulations of circumbinary disks predicts that MBH binaries
can excite periodic enhancements of the mass accretion rate that could
translate into periodic luminosity enhancements, not only at the
orbital period, but also on longer and shorter timescales. These
periodic modulations are on timescales corresponding to $\approx$1/2
and 1 times the binary's orbital period. For high BH mass ratios
($q\equiv M_1/M_2 \gtrsim 0.3$), the cavity is lopsided, leading to the
formation of a hotspot in the accretion disk. The strongest modulation
in the accretion rate in this case is observed at the orbital period
of the overdense region, a few ($\sim$3-8) times the orbital period of
the binary.

In addition, both BHs should have their own photospheres in X-ray and
possibly also in optical bands~\citep{Haiman2017}, because the
empirical sizes (from microlensing and variability studies) show that
the X-ray emitting regions of quasars have sizes of a few $R_s$,
whereas the separation of massive binaries entering the detector
frequency range is $\approx 100 R_s$.  Relativistic Doppler
modulations and lensing effects will then inevitably imprint periodic
variability in the EM light-curve at the several percent level,
tracking the phase of the orbital motion, and serving as a template
for the GW inspiral waveform.  The Doppler-induced variability
amplitude will increase over time, as in the GW
chirp~\citep{Haiman2017,Schnittman+2018}, while self-lensing for near
edge-on binaries would imprint additional characteristic, periodically
recurring spikes on the EM light-curve~\cite{D'Orazio+2018}.

\subsubsection{Co-evolution of massive black holes and their host galaxies}
Nuclear MBHs correlate with many properties of their host galaxies,
suggesting that MBHs and galaxies co-evolve over cosmic
time. However, the nature of this co-evolution and the physics
responsible for it is not yet understood~\cite{Kormendy2013}.  If EM
observations can identify unique host galaxies of the MBH binaries
detected in GWs then this will directly provide the relation between
(merging) MBHs and their host galaxies as a function of redshift, as
well as luminosity and other parameters.  The GW data will yield
precise and reliable estimates of the masses (as well as orbital
parameters and spins) of the MBHs, which will not be available from
EM observations alone.

\subsubsection{What can neutrinos tell us?}
Correlation between a GW detector with high angular resolution and cosmic neutrino detectors can yield new information on energetic particle acceleration, binary BH mergers and the neutrino mass. In particular, the origin of the observed diffuse neutrino flux remains unknown. The recent detection of a high-energy neutrino in the direction of the gamma-ray emitting blazar TXS 0506+056 \cite{147} strongly suggests that blazars, and AGNs in general, may be a source of high-energy neutrinos. Searches of coincident neutrino-GW emission have not led to a successful detection yet \cite{PhysRevD.93.122010}, and obtaining high angular resolution for GW observations can significantly improve the prospects of a neutrino-GW observation. Even if neutrino flux and GW signals cannot be correlated in time or position, high angular resolution observation of GW sources may allow for population studies from which a diffuse neutrino flux may be derived that might explain the observed neutrino spectrum.



Recent observation of a high-energy neutrino in the direction of a blazar is consistent with a hadronic neutrino emission from a relativistic jet emitting beamed gamma-rays towards Earth, which can occur for large BH masses \cite{147}. Moreover, modelling suggests that the coalescence of two MBHs with aligned jets pointing towards Earth to get maximum Doppler boost, would provide the required shock acceleration just before impact to give an observable peak in gamma and neutrino flux \cite{refId0}.
The directional properties of the GW detection would be valuable in identifying the neutrino source while the time relation between the GW waveform and gamma ray and neutrino arrival will help elucidate the history of the jet mergers or even the neutrino mass. Neutrinos would take up to $\sim 1000$ sec longer to arrive than photons. If much better knowledge of the neutrino mass is available pre-launch, the difference in flight time between photons and neutrinos is an additional diagnostic.

Another prime candidate for neutrino-GW co-observation are tidal disruption events (TDE). A TDE may occur when a main sequence star or white dwarf is disrupted when it passes a MBH within a critical distance in a highly eccentric orbit. TDE candidate events have been observed in the optical/UV/X-ray spectrum \cite{Auchettl_2017, KOMOSSA2015148}. Single and quiet MBHs may be detected through high luminosity flares during a TDE. These flares may form a neutrino source, but perhaps more interesting events occur when the disrupted star is found in a binary system of MBHs (or IMBHs). These phenomena, which are expected to occur at a much higher rate, would provide a source of neutrinos in the high energy or even ultra-high energy (TeV - EeV) range \cite{PhysRevD.95.123001}.

\subsection{Tests of General Relativity} 
\label{sec:TGR}

%
 Theories aiming to explain the late-time acceleration of the Universe typically introduce new fields beyond the Standard Model of particle physics, whose interactions are predominantly gravitational. Their impact on the dynamics of GWs has observable effects which can be broadly divided into two distinct kinds: i) modifications of the GW waveform as a result of the modified dynamics of the merger, and ii) subsequent effects due to propagation through a (modified) cosmological environment. Constraints on gravity from the production and/or propagation of GWs are complementary to those obtained from studies of the large-scale structure of the Universe. Whilst the propagation of GWs over cosmological distances can yield relatively clean tests of modifications to GR, the dynamics of merging events is substantially more difficult to analyse. For example, in some systems, `screening mechanisms' act to suppress the effects of additional degrees of freedom near to GW sources \cite{Babichev:2013usa,Burrage:2017qrf} , potentially resulting in GW waveforms at emission that are indistinguishable from those of GR.

Below we shall first discuss the distinctive signals of alternative gravity models upon GW propagation over cosmological distances; then we will comment on some of their other effects on the emitted waveform. We shall find a common pattern that identification of host galaxies or electromagnetic counterparts is key to many of these tests, and hence a high-angular resolution GW detector will facilitate this science.

\subsubsection{Tests using cosmological propagation of GWs}
If we assume that GWs travel away from the source as plane waves, their propagation on a homogeneous and isotropic background can be generically described through the following equation (in conformal time $t$) \cite{Saltas:2014dha, Nishizawa:2017nef, Ezquiaga:2018btd, Bettoni:2016mij}:
\begin{equation}
h''_{ij}+ \big[2+\nu(t) \big]H(t) h'_{ij} + \left[1 + \alpha_{\mathrm T}(t)\right] k^2h_{ij}+a^2\mu^{2}h_{ij}=a^2\Gamma(t)\gamma_{ij}(t)\,,   \label{eq:GWeq}
\end{equation}
with the case of GR corresponding to $\nu(t), \alpha(t), \mu^{2}, \Gamma(t) = 0$. This simple equation captures essential features of generic theories which modify GR through a new dynamical scalar, vector or tensor field.  In particular, the modification to the friction term, through $\nu(t) \equiv H(t)^{-1}d \ln M(t)^2/dt$, where $H(t)$ is the Hubble parameter, signifies a running of the Planck mass $M$ with time. This is a typical feature of conformal couplings between scalar fields and curvature, as predicted within general scalar-tensor theories \cite{Horndeski:1974wa, horndeski, Langlois:2015cwa,Crisostomi:2016czh,BenAchour:2016fzp, Langlois:2017mxy}). In addition, the existence of non-trivial interactions between the metric and the new field modifying the causal structure of gravitons will affect the propagation speed $\alpha_{\mathrm T}(t) \neq 0$. 
Finally, a new tensor field interacting with the spacetime metric acts as a source for the GW propagation equation through the coupling $\Gamma(t)$. This effect occurs in massive gravity and bigravity \cite{2014LRR....17....7D}, and will lead to a modified dispersion relation \cite{Hassan:2011zd} and GW oscillations \cite{Max:2017flc} (see sections \ref{speed} and \ref{sec:gw_oscillations}).

The $\alpha$ functions appearing in (\ref{eq:GWeq}) can be mapped onto model-independent cosmological observables, e.g., the gravitational slip parameter $\eta$ \cite{Saltas:2014dha}. This connection establishes a way to combine GW measurements with future high-precision cosmological surveys such as the Euclid satellite and the LSST. 

\smallskip

As a representative, well-studied class of models, let us now discuss the theoretical structure of scalar-tensor theories in more detail. We focus on theories in the Horndeski class \cite{Horndeski:1974wa,Deffayet:2011gz,Kobayashi:2011nu}, characterized by second order equations of motion, and including Brans-Dicke, covariant Galileons and many others. For brevity, we discuss here theories with $\alpha_T=0$ (in order to be in agreement with constraints from GW170817, but see the comment at the end of this section). They are described by the action:
 \begin{equation}\label{h-action}
  S_{\rm Horndeski}^{c_T=1}\,=\,\int d^4 x\,\sqrt{-g}\,\left[ G_{2}(\phi, \,X) \,+G_3(\phi, \,X) \,\left(\nabla_\mu \partial^\mu \phi\right)+G_{4} (\phi)\,R\right]\,,
 \end{equation}
 with $G_{2,3,4}$ arbitrary functions, $\phi$ the scalar field, and $X\,\equiv\,\nabla_\mu \phi \nabla^\mu \phi$. These theories predict vanishing $\mu$ and $\Gamma$ in equation (\ref{eq:GWeq}), but the function $G_4(\phi)$ leads to a non-vanishing form for the function $\nu(t)$ in equation (\ref{eq:GWeq}), which can be probed via the GW luminosity distance \cite{Deffayet:2007kf,Camera:2013xfa,Nishizawa:2017nef,Belgacem:2017ihm,Arai:2017hxj,Nishizawa:2019rra,Belgacem:2018lbp} (see section \ref{sec-ld}).
The scalar derivative  self-interactions in equation (\ref{h-action}), controlled by the functions $G_{2,3}$, automatically provide Horndeski  theories with a Vainsthein screening mechanism, allowing for consistency with Solar System tests (see \cite{Babichev:2013usa} for a review). There have been only a few studies so far analyzing the (complex) problem of GW emission from merging binaries in theories which screen \cite{Dar:2018dra,2016PhRvL.116f1101B}.

A  final comment on our assumption $\alpha_T=0$ (equivalently, $c_T=1$). The scalar-tensor theories described above have a 
low UV cut-off, of order $\Lambda\,=\,\left( H_0^2\,M_{\rm Pl}\right)^{1/3}\,\sim\,10^{2}$ Hz. This value of $\Lambda$ is precisely within the frequency range probed by the LIGO-Virgo detectors. Hence it may be that new physics near the cut-off changes theoretical predictions,  possibly making the restriction to $c_T=1$ unnecessary at GW frequencies probed by ground based detectors \cite{deRham:2018red}. On the other hand, space-based
experiments like LISA will probe GW frequencies in ranges well below the cut-off, and will be able to test theoretical predictions of scalar tensor theories well within the range of validity of eq.(\ref{h-action}). These and related theoretical considerations suggest that there may be a dependence on frequency of some of the time-dependent parameters of equation (\ref{eq:GWeq}). We shall discuss this possibility further in the following section.

\subsubsection{GW propagation speed and dispersion tests}
\label{speed}
Modified gravity theories can alter the dispersion relation of GWs; we can write these changes in a parameterised format as:
\begin{equation}
    E^2 = p^2c^2 + A_q p^q c^q
\end{equation}
where $p$ is the magnitude of the three-momentum, the index $q$ ({not} to be confused with a spacetime index) runs over possible power law dependencies of the MG corrections ($q=0, 1, 2, 3, \ldots$), and $A_q$ is an amplitude with the appropriate dimensions for consistency (equivalent to $m^{2-q} c^{4-2q}$). For $q=2$ these corrections reduce to a straightforward modification of the GW speed, as present in eq.(\ref{eq:GWeq}).
Other values of $q$ arise in, for example, bigravity ($q=0$ represents a massive graviton) \cite{2014LRR....17....7D} , doubly special relativity ($q=3$) \cite{2002Natur.418...34A} and Hor\u{a}va-Liftschitz gravity ($q=4$) \cite{2009PhRvD..79h4008H}. Constraints on the set of parameters $A_q$ from the LIGO-Virgo detectors are presented in \cite{Abbott:2018lct}.

For the case $q=2$, we can write the deviation of $c_T$ from the speed of light as:
\begin{equation} 
c_T^2=c^2[1+\alpha_T(z,f)]
\end{equation}
The notation here is inherited from Horndeski theory, but promoting $\alpha_T$ to a function of both frequency and redshift, as discussed at the end of the previous section. By measuring the delay in arrival time between GWs from a merger event and the photons associated to its electromagnetic counterpart, we can place a bound on $\alpha_T$ \cite{Lombriser:2015sxa,Bettoni:2016mij}. In some cases one may need to account for a sizeable delay (up to a few hundred seconds) between the emission of GWs and photons.

The detection of a gamma-ray counterpart to event GW170817, arriving 1.74 seconds after the GW merger, yielded the exceptionally strong bound $|\alpha_T|\leq 10^{-15}$ at $z\sim 0$ and $f\sim 250$ Hz \cite{Abbott:2018lct,Baker:2017hug,PhysRevLett.119.251304,Sakstein:2017xjx,2017PhRvL.119y1302C} (assuming no intrinsic emission delay between GWs and photons). This is comparable to the one-sided bound obtained at very high energies from the lack of observed gravi-Cherenkov radiation by cosmic rays \cite{2001JHEP...09..023M}. As yet, no such comparable bounds have been obtained at low (millihertz) frequencies or higher redshifts. LISA will offer the first opportunity to do this \cite{Bettoni:2016mij}, generally resolving the sky localisation of massive black hole binaries to 10 deg$^2$ hours or weeks prior to merger, and down to 1 deg$^2$ for low-redshift sources \cite{2017arXiv170200786A,2008ApJ...677.1184L,2008ApJ...684..870K} \footnote{Sky localization for LISA can improve significantly in the final moments before merger, but this extra information may often come too late to slew additional electromagnetic facilities. Likewise adding ringdown and merger information can boost localization \textit{a posteriori} \cite{2016JCAP...04..002T}.}. Although wide-field survey telescopes like LSST \cite{2009arXiv0912.0201L} can cover a field of view of 10 deg$^2$, due to the high galaxy density, multiple varying sources in the field could confuse the host identification process. Furthermore, other observatories have considerably smaller fields of view. For the counterpart of GW170817, having a multitude of observatories at different wavelengths was crucial to pinning down properties of the source \cite{2017ApJ...848L..13A}. 

Therefore, identifying the potential host galaxies of a larger number of GW sources at an earlier stage would strongly benefit tests of gravity. A low frequency GW detector with roughly arcminute angular resolution can assist with this,  and hence boost the likelihood of observing a prompt counterpart across the electromagnetic spectrum. Just as the LIGO-band constraints on $\alpha_T$ have had a profound effect on theories of modified gravity, millihertz bounds on the propagation speed and dispersion relation of GWs will enable us to rule out or validate other remaining families of models. Identification of hosts at higher redshifts, where LISA's localisation worsens, will also confirm that the bounds from GW170817 hold true at all redshifts. 

In addition, as argued above, a high-resolution GW detector will be able to localise a massive binary BH event on the sky well in advance of the merger (up to several months).  This will allow to point an X-ray or UV telescope at the source, and monitor it during its final O(1000) inspiral cycles.  As discussed in Section \ref{sec:agn}, both BH components will likely have their own separate compact photospheres during these final months, and therefore the binary is expected to be producing a bright EM chirp signal, in X-ray and possibly also in
UV and optical bands.  Relativistic Doppler modulations and lensing effects will inevitably imprint periodic variability on the EM light-curve, which will track the phase of the GWs, since
both the EM and GW chirp are caused by the same orbital motion, and will therefore serve as a template for the GW inspiral
waveform~\cite{Haiman2017}.  Note that these kinematic effects can be utilised without modeling the EM source engine itself (this modeling is otherwise a major uncertainty in comparing arrival times of photons vs gravitons, and, in the case of GW170817, weakened the limit on the graviton mass by orders of magnitude). A comparison of the phases of the GW and EM chirp signals will help break degeneracies between system parameters, and probe a fractional difference difference
$\Delta v$ in the propagation speed of photons and gravitons as low as $\Delta v/c \approx 10^{-17}$ at LISA-like sensitivities.

\subsubsection{GW luminosity distance tests.}
\label{sec-ld}
Another effect of modified gravity on cosmologically propagating GWs arises due to the modified damping term $\nu$ in eq.(\ref{eq:GWeq}). In GR, the amplitude of a GW is inversely proportional to the luminosity distance of the source, $h_{GR}\propto 1/d_L(z)$ (modulo factors involving the mass, frequency and inclination angle of the source). Modifications to the damping term alter this relation to be:
\begin{equation}
h_{MG}\propto \frac{e^{-D(z)}}{d_L(z)}\equiv \frac{1}{d_{GW}(z)}\quad\quad\quad\quad \mathrm{where}  \quad\quad   \quad\quad D(z)=\frac{1}{2}\int_0^z\frac{\nu(\tilde{z})}{(1+\tilde{z})}d\tilde{z}\,.
\end{equation}
The effective luminosity distance of the GW source, $d_{GW}$, now differs from the luminosity distance of its electromagnetic counterpart, $d_L$. Like the anomalous speed tests of section~\ref{speed}, this modified damping test requires the identification of the source redshift, such that $d_{GW}$ and $d_L$ can be independently measured and bounds placed on $\nu$. This test can be directly performed if an electromagnetic counterpart is identified, but it can also be statistically applied by cross-correlating GW events with galaxy catalogs.
LISA can perform this test with multi-messenger GW events \cite{belgacem2019}; however, the need for a clear EM counterpart limits the number of standard sirens available and uncertainties in the identification translate into larger error bars in the luminosity distance. By using a high angular resolution detector to narrow down the host galaxy candidates, more events can be confidently assigned a redshift, thus increasing the precision of our constraints on dark energy and modified gravity \cite{Ezquiaga:2018btd,belgacem2019} (see also Section \ref{sec:cosmology}). Moreover, an improved sky localization will help in breaking the distance-inclination degeneracy, which accounts for a large fraction of the error in $d_{GW}$.

\subsubsection{GW oscillations}
\label{sec:gw_oscillations}
Another propagation effect of interest is the phenomenon of GW oscillation. This can arise in gravity theories with additional fields, if the interaction and propagation eigenstates differ. For example, in massive bigravity there is an interaction with the second tensor field \cite{Max:2017flc} that behaves similarly to neutrino flavour oscillations. 
Oscillations modulate the strain as a function of the distance to the source and the GW frequency, leading to an interference pattern that can be detected/constrained by GW observations alone. 
The interference pattern is lost \cite{Max:2017kdc} when the wavepackets of the two eigenstates do not overlap, if the graviton mass and distance to the source are large. 
This decoherence limit can be tested by using either multiple signals (two GW packets arriving from the same source) or standard sirens, through the decrease in GW amplitude. In both cases an identification of the counterpart through improved sky localization is highly desirable, either to conclude that two signals have the same origin, or to provide a redshift for the luminosity distance.

%
\subsubsection{Additional polarizations}
\label{pol}
Multiple detectors are necessary to fully measure all possible gravitational polarizations for a given signal. 
Metric theories of gravity predict at most six gravitational polarizations: two tensors, two vector and two scalar helicities. Separation of all the components is an under-determined problem since only six observables exist (the independent components of the Riemann tensor $R_{0i0j}$), yet eight unknowns need to be determined: six polarization amplitudes, plus two angles for the GW incidence direction \cite{Gair:2012nm}. (Note that different polarizations have different sensitivity patterns \cite{Nishizawa:2009bf,Tinto:2010hz}, and that for differential-arm detectors the two scalar modes are degenerate, so only five polarisations are measurable \cite{2019arXiv190304467T}.) Only by incorporating the time-of-flight information between the different detectors is it possible to optimally characterize all possible components of a signal.
Further improvements might be achieved by optimizing the location and orientation of the two detectors (e.g., coplanar vs. orthogonal).

In the case of ground-based detectors, the advantage of using precise localization of the GW events to improve the polarization tests was demonstrated explicitly using the BNS merger event GW\,170817. When the position of the electromagnetically identified counterpart of GW\,170817 was used, the Bayes factors in the polarization tests improved by almost $\sim20$ orders of magnitude~\cite{Abbott:2018lct}. Roughly speaking, the position information breaks a large portion of degeneracy by reducing the parameter space contained in the pattern functions, which is expected to be true for space-based detectors with high angular resolution as well.

\subsubsection{Modified waveforms }
\label{sec:waveform}
Modifications to GR that persist in the strong-field regime -- i.e., are not screened by a host galaxy as discussed above -- can alter the waveform emitted by a binary. To avoid inefficiently testing theories on a case-by-case basis, the authors of \cite{2009PhRvD..80l2003Y} have developed a model-independent formalism called the Parameterized Post-Einsteinian framework (PPE). PPE provides a `template' modified waveform onto which many different theories of gravity can be mapped, see table 1 of \cite{2011PhRvD..84f2003C}. A simplified limit of PPE models the waveform as:
\begin{equation}
h_{MG} = h_{GR}\left[1+\alpha u^a\right]e^{-\beta u^b}    
\end{equation}
where $h_{GR}$ is the GR inspiral waveform, $u =\pi {\cal M}f$ ($\cal M$ is the chirp mass), and the modifications are encapsulated in the four parameters $\{\alpha, \beta, a, b\}$. Forecasts for how well these parameters can be constrained using LIGO and LISA data are given in \cite{2011PhRvD..84f2003C}.

Improved identification of GW host galaxies adds an interesting new layer to this investigation, as it allows the onset of screening to be studied. For example, if we were to hypothetically detect non-GR values of $\{\alpha, \beta, a, b\}$ in some systems, we could assess whether these systems all live in (say) low-mass galaxies. This would fit with a model in which screening efficiency scales with the depth of the gravitational potential, as occurs in chameleon screening \cite{Burrage:2017qrf}. In this way, a high-angular resolution GW detector could enable a direct linking of strong-field and cosmological tests of modified gravity.

\subsubsection{Dipole radiation and multiband GW detection}
Dipole radiation exists in a variety of alternative gravity theories, the
canonical example being Jordan-Fierz-Brans-Dicke (JFBD)
theory~\cite{Will:2018bme} which is now tightly constrained by Solar System observations~\cite{1994PhRvD..50.6058W}. However, in extended variants of JFBD theories, the possibility of a large dipole radiation component still exists, especially when strongly self-gravitating
neutron stars undergo the so-called ``spontaneous scalarization'' and
``dynamical scalarization'' phenomena~\cite{1993PhRvL..70.2220D, 2013PhRvD..87h1506B,
2017PhRvX...7d1025S}. 
Recently, black holes are also found to be scalarized in
some classes of scalar-tensor theories~\cite{2016PhRvD..93b4010Y}.  Therefore,
all three kinds of compact binaries --- BNSs, BBHs, and NS-BH binaries --- can in principle possess extra dipole emission with respect to the quadrupole radiation predicted in GR.

Because dipole radiation enters a binary system at the
$-1$ post-Newtonian order, it can be constrained by low-frequency observations from space-based detectors~\cite{2017PhRvD..96h4039C}, or even by multi-band GW observations from both space-based and ground-based detectors~\cite{2016PhRvL.116w1102S, 2016PhRvL.116x1104B}. A high angular resolution detector would assist with host galaxy identification of the source at early stages, and hence information about source environments can be factored into our bounds on dipole radiation (similar to the discussion in the second paragraph of section~\ref{sec:waveform}). Dipole radiation is usually accompanied by extra polarization mode(s), for example, an extra scalar mode in scalar-tensor gravity~\cite{1994PhRvD..50.6058W}. As discussed in section~\ref{pol}, high angular resolution helps to distinguish between different polarization modes, thus also boosts sensitivity to additional channels of gravitational radiation.\\

\vspace{5mm}

\section{Summary}
\label{sec:summary}
Gravitational wave astronomy is an observational discipline in its infancy, but it has already provided many new insights into our Universe. Many more exciting discoveries lie ahead, but realisation of the full scientific potential will require major improvements to our observational capabilities, in particular angular resolution, which is the focus of this White Paper. The accuracy of source localisation by the current gravitational wave detectors and even those currently under development is far inferior to that of electromagnetic telescopes, ranging from O(10) to O(1) degrees in the best cases. High angular resolution is nevertheless paramount for multi-messenger observations, identification of the host galaxies of gravitational wave sources and observation of the stochastic background and its anisotropies. \\

High angular resolution can be obtained by increasing the effective aperture of the gravitational wave telescope. As outlined in Section \ref{sec:strawman}, this can be achieved by having more than one detector with a baseline of $\sim$AU. This setup would most probably necessitate a collaboration with an international partner, but is in principle possible for an L-class mission on the Voyage 2050 timescale.\\

High angular resolution gravitational wave observations will provide us with unique tools to answer fundamental open questions in cosmology and astrophysics. As discussed in Section \ref{sec:cosmology}, arcmin resolution of astrometric positions will ensure the identification of the host galaxy for the vast majority of detected sources (with a virtually guaranteed electromagnetic counterpart for all the BNS detections), increasing the number of standard sirens to $10^5-10^6$ and allowing precision measurement of cosmological parameters. Moreover, since the population of gravitational wave sources is a biased tracer of the underlying dark matter distribution, cross-correlations with galaxy surveys will provide a new handle on cosmological structure formation. Much is to be learnt from the as yet undetected stochastic gravitational wave background, which can be produced by astrophysical sources, as well as by processes in the early Universe, such as inflation and phase transitions. As we showed in Section \ref{sec:stochastic}, a gravitational wave telescope with a $\sim$AU effective baseline is the optimal instrument for resolving the angular features of the stochastic background. High resolution gravitational wave observations will also vastly facilitate multi-messenger astronomy. As discussed in Section \ref{sec:MBH}, the ability to localise a massive black hole binary months before the merger will allow the observation of features in the electromagnetic spectrum related to the structure of the accretion disk. Moreover, coincident neutrino-gravitational wave observations may elucidate the origin of high-energy neutrinos and the diffuse neutrino flux, at present unknown. Finally, the ability to localise gravitational wave sources is also fundamental for studying modified gravity and physics beyond the Standard Model by testing the propagation speed, dispersion and polarization of gravitational waves, as described in Section \ref{sec:TGR}.

%

\newpage
\renewcommand*{\bibfont}{\normalsize}

\printbibliography[heading=subbibliography]
\newpage
\thispagestyle{empty} 
\section*{Proposing team}
\onehalfspacing
\large
John Baker (Goddard Space Flight Centre, US)\\
Tessa Baker (Queen Mary University of London, UK) \\
Carmelita Carbone (INAF - Institute of Space Astrophysics and Cosmic Physics, Milano, Italy)\\
Giuseppe Congedo (Institute for Astronomy, University of Edinburgh, UK)\\
Carlo Contaldi (Imperial College London, UK)\\
Irina Dvorkin (Albert-Einstein-Institute, Potsdam, Germany)\\
Jonathan Gair (Albert-Einstein-Institute, Potsdam, Germany)\\
Zoltan Haiman (Columbia University, US)\\
David F. Mota (University of Oslo, Norway)\\
Arianna Renzini (Imperial College London, UK)\\
Ernst-Jan Buis (TNO, Delft, the Netherlands) \\
Giulia Cusin (University of Oxford, UK)\\
Jose Maria Ezquiaga (Universidad Aut\'{o}noma de Madrid, Spain)\\
Guido Mueller (University of Florida, US)\\
Mauro Pieroni (IFT, Universidad Aut\'{o}noma de Madrid, Spain)\\
John Quenby (Imperial College London, UK) \\
Angelo Ricciardone (INFN, Sezione di Padova, Italy)\\
Ippocratis D. Saltas (CEICO, Institute of Physics of the Czech Academy of Sciences, Prague, Czechia) \\
Lijing Shao (Kavli Institute for Astronomy and Astrophysics, Peking University, Beijing, China)\\
Nicola Tamanini (Albert-Einstein-Institute, Potsdam, Germany)\\
Gianmassimo Tasinato (Swansea University, Swansea, UK)\\
Miguel Zumalac\'{a}rregui (University of California at Berkeley, US)
\end{document}